\documentclass[twocolumn,floatfix,showpacs,aps,nofootinbib]{revtex4}
\usepackage{amssymb,amsmath}
\usepackage{graphicx,float}
\usepackage{indentfirst}
\newcommand{\be}{\begin{equation}}
\newcommand{\ee}{\end{equation}}

\newcommand{\ba}{\begin{eqnarray}}
\newcommand{\ea}{\end{eqnarray}}

\newcommand{\el}{^}
\begin{document}

\title{Unfolding Physics from the Algebraic Classification of Spinor Fields}

\author{J. M. Hoff da Silva}
\email{hoff@feg.unesp.br;hoff@ift.unesp.br}
\affiliation{Departamento de F\1sica e Qu\1mica, Universidade
Estadual Paulista, Av. Dr. Ariberto Pereira da Cunha, 333,
Guaratinguet\'a, SP, Brazil.}
\author{Rold\~ao da Rocha}
\email{roldao.rocha@ufabc.edu.br}
\affiliation{Centro de Matem\'atica, Computa\c c\~ao e Cogni\c c\~ao, Universidade Federal do ABC 09210-170, Santo Andr\'e, SP, Brazil.}


\begin{abstract}
After  reviewing the Lounesto spinor field classification, according to the bilinear covariants associated to a spinor field, we call attention and unravel some prominent features involving  unexpected properties about spinor fields under such classification. In particular, we pithily focus on the new aspects --- as well as current concrete possibilities. They mainly arise when we deal with some non-standard spinor fields concerning, in particular, their applications in physics.
\end{abstract}

\maketitle

\flushbottom

\section{Introduction}

From the classical point of view, the definition of spinors is based upon irreducible  representations of the group $Spin_{+}(p,q)$, where $p+q = n$ is the spacetime  dimension. Due to the immediate physical  interest,  mainly the Minkowski spacetime $\mathbb{R}^{1,3}$ has being regarded since the 1920's.
On the another hand, the representation space associated to  a
irreducible regular representation in a Clifford algebra
 is a minimal left ideal. Its elements are the so-called algebraic spinors.
Another possible definition of a spinor, which is denominated
{operatorial}, can be introduced from another representation
-- distinct of the regular representation -- of a Clifford algebra, using the representation space associated to the even  subalgebra.  This definition  is
equivalent to the classical and algebraic ones, in particular in the cases of great interest for physical applications.
The classical definition of spinor is the customary approach in several superb textbooks in physics, e. g., \cite{wein}.  There is no damage in asserting that, in Minkowski spacetime, classical spinors are irreducible  representations of the Lorentz group $Spin_{+}(1,3) \simeq SL(2, \mathbb{C})$. Notwithstanding,  this paradigm  severely restricts the analysis to the usual Dirac, Weyl, and Majorana spinors.

A new possibility involving the spinor fields classification  was introduced by Lounesto \cite{LOUN2}, as a palpable paradigm shift. It is based upon the bilinear covariants and their underlying multivector  structure. In particular, this classification evinces the existence of a new type of spinor field, the so called flag-dipole spinor fields. Furthermore, it additionally  presents another class  of spinor fields (the flagpoles) that accommodates Elko spinor fields, which are prime candidates to the dark matter description \cite{AHL}.
They  generalize Majorana spinor fields. As it is well known, any spin-half spinor field, that potentially describes the dark matter, respects the symmetries of the Poincar\'e group in the sense of Weinberg, if it is an element of a standard Wigner class of representations of the Poincar\'e group. As it will be reported, Elko spinor fields do not belong to the standard Wigner class. Among a significant amount  of unexpected and interesting properties, it was recently demonstrated that the topological exotic spacetime structure
can be probed uniquely by Elko spinor fields: they are, hence, suitable to investigate the eventual non-trivial topology of the universe \cite{RBJ}. By such exoticness,  dynamical constraints converted into a dark spinor mass generation mechanism, with the encrypted VSR symmetries holding as well.

The aim of this work is to report some of the recent advances in this field of research, calling special attention to the interesting features associated to the new spinor fields appearing in the Lounesto's classification. In order to accomplish that, we organize this work as follows: in the next Section we review the formal and necessary aspects regarding the Lounesto spinor classification. In Sec. III, we explore some of the odd and captivating aspects associated to Elko and flag-dipole spinor fields. In the final Section we conclude.

\section{Classifying spinor fields}

We start this Section reviewing some indispensable preliminary concepts. For a deeper approach see, e. g., \cite{DET}. Consider the tensor algebra $T(V)=\bigoplus_{i=0}^{\infty} T\el{i}(V)$, where $V$ is a finite $n$-dimensional real vector space. Henceforth $V$
is regarded as being  the tangent space on a point on a manifold. Let $\Lambda\el{k}(V)$ denote the antisymmetric $k$-tensors space, indeed the $k$-forms vector space. In this way $\Lambda(V)=\bigoplus_{k=0}\el{n}\Lambda\el{k}(V)$ is the space of the differential forms over $V$.
   For any $\psi \in \Lambda(V)$,
    the  reversion is defined by $\tilde{\psi}=(-1)\el{[k/2]}\psi$ (the integer part of $m$ is denoted by $[m]$), which is an antiautomorphism in $\Lambda(V)$. Moreover, $\hat{\psi}=(-1)\el{k}\psi$ denotes the graded involution, also called main automorphism.  It is possible to use the metric $g:V^*\times V^*\rightarrow \mathbb{R}$ extended to the $k$-forms space, in order to define the left and right contractions. Hence, for $\psi=\wedge_{i=1}\el{p}{\bf u}\el{i}\equiv{\bf{u}}\el{1}\wedge\cdots \wedge {\bf u}\el{p}$ and $\phi=\wedge_{j=1}\el{r}{\bf v}\el{r}$, with ${\bf u}\el{i},{\bf v}\el{j}\in V\el{*}$, the extension of $g$ to $\Lambda(V)$ reads $ g(\psi,\phi)=det(g({\bf u}\el{i},{\bf v}\el{j}))$ for $p=r$, and zero otherwise. Now one defines the left contractionby \be g(\psi\lrcorner\varphi,\chi)=g(\varphi,\tilde{\psi}\wedge\chi)\label{left},\qquad\text{ for $\psi$, $\varphi$, $\chi$ $\in \Lambda{(V)}$.} \ee For ${\bf v} \in V$, the Leibniz rule for the contraction is \be {\bf v}\lrcorner(\psi\wedge \varphi)=({\bf v}\lrcorner \psi)\wedge\varphi+\hat{\psi}\wedge({\bf v}\lrcorner \varphi)  \ee respectively. The Clifford product between  ${\bf v} \in V$ and $\chi \in \Lambda(V)$ is ${\bf v}\chi={\bf w}\wedge\chi+{\bf v}\lrcorner \chi$ and the pair  $(\Lambda(V),g)$, endowed with the Clifford product, is denoted by $Cl(V,g)$ ($Cl_{p,q}$ is a notation that shall be reserved to the Clifford algebra when $V \simeq\mathbb{R}\el{p,q}$).

In order to properly revisit the bilinear covariants let us fix the gamma matrices notation. All the formalism in representation independent, and hence we use hereon the Weyl (or chiral) representation of $\gamma\el{\mu}$:
$
\gamma_{0}=\gamma^{0}=%
\scriptsize{\begin{pmatrix}
{\mathbb{O}} & {\mathbb{I}} \\
{\mathbb{I}} & {\mathbb{O}}%
\end{pmatrix}
,\,\gamma_{k}=-\gamma^{k}=%
\begin{pmatrix}
{\mathbb{O}} & \sigma_{k} \\
-\sigma_{k} & {\mathbb{O}}%
\end{pmatrix}}
$, where
$
{\scriptsize{{\mathbb{I}}=
\begin{pmatrix}
1 & 0 \\
0 & 1%
\end{pmatrix}
,\; {\mathbb{O}}=%
\begin{pmatrix}
0 & 0 \\
0 & 0%
\end{pmatrix}}}
$ and the $\sigma_i$ are the Pauli matrices. Moreover $
\gamma^{5}=i\gamma^{0}\gamma^{1}\gamma^{2}\gamma^{3}$. All the spinor fields in this work are placed in the Minkowski spacetime  $(M\simeq \mathbb{R}\el{1,3}, \eta, D, \tau,\uparrow)$, where $\eta={\rm diag}(1,-1,-1,-1)$ is a metric which has a compatible (Levi-Civita) connection $D$ associated. Besides, $M$ has spacetime orientation induced by the volume element $\tau$ as well as time orientation  denoted by $\uparrow$. We denote by $\{x\el{\mu}\}$  global coordinates, in terms of which an inertial frame ---  a section of the frame bundle ${\bf P}_{SO_{1,3}}(M)$
--- reads ${\bf e}_{\mu}=\partial/\partial x\el{\mu}$.

At this point we recall that classical spinor fields are sections of the vector bundle ${\bf P}_{Spin_{1,3}}\times \mathbb{C}\el{2}$, where the specific representation of $SL(2,\mathbb{C}) \simeq Spin_{1,3}$ in $\mathbb{C}\el{2}$ is implicit. In this framework, the bilinear covariants associated to a spinor field $\psi \in {\bf P}_{Spin_{1,3}}\times \mathbb{C}\el{2}$ are sections of $\Lambda(TM)$ into the Clifford bundle of multiform fields, given by \begin{align}
\sigma & =\psi ^{\dagger }\gamma _{0}\psi ,\qquad \mathbf{J}=J_{\mu }\theta
^{\mu }=\psi ^{\dagger }\gamma _{0}\gamma _{\mu }\psi \theta ^{\mu }\nonumber\\
\mathbf{S}&=S_{\mu \nu }\theta ^{\mu \nu }=\frac{1}{2}\psi ^{\dagger }\gamma
_{0}i\gamma _{\mu \nu }\psi \theta ^{\mu }\wedge \theta ^{\nu }  \notag \\
\mathbf{K}& =K_{\mu }\theta ^{\mu }=\psi ^{\dagger }\gamma _{0}i\gamma
_{0123}\gamma _{\mu }\psi \theta ^{\mu },\;\;\omega =-\psi ^{\dagger
}\gamma _{0}\gamma _{0123}\psi  \label{fierz}
\end{align}%
where $\{\theta\el{\mu}\}$ is the dual basis of $\{{\bf e}_{\mu}\}$. The bilinear covariants obey quadratic equations, the so called Fierz-Pauli-Kofink identities \cite{LOUN2}
\begin{eqnarray}
\mathbf{J}\llcorner\mathbf{K}&=&0,\qquad\mathbf{J}^{2}=\omega^{2}+\sigma^{2}\nonumber\\\mathbf{J}\wedge\mathbf{K}%
&=&-(\omega+\sigma\gamma_{0123})\mathbf{S}, \qquad
\mathbf{K}^{2}=-\mathbf{J}%
^{2}, \label{fi}
\end{eqnarray} which are particularly interesting in what follows. The Fierz aggregate $Z$ is defined by \be Z=\sigma+\mathbf{J}+i\mathbf{S}-i\gamma_{0123}\mathbf{K}+\gamma_{0123}\omega.\label{ZZZ} \ee Eqs. (\ref{fierz}) may be recast in terms of $Z$, yielding
\begin{align}
Z\el{2} & =4\sigma Z ,\quad Z\gamma_{\mu}Z=4J_{\mu}Z,\quad
Zi\gamma_{\mu\nu}Z=4S_{\mu\nu}Z,  \notag \\
\quad Z\gamma_{0123}Z&=-4\omega Z,\quad Zi\gamma_{0123}\gamma_{\mu}Z  =4K_{\mu}Z \label{Zfierz}
\end{align}
Therefore, it is possible to categorize different spinor fields by different $Z$'s, or similarly by distinct bilinear covariants. The Lounesto spinor field classification provides the following spinor field
classes \cite{LOUN2}:
\begin{itemize}
\item[1)] $\sigma\neq0,\;\;\; \omega\neq0$\qquad4) $\sigma= 0 = \omega, \;\;\;\mathbf{K}\neq 0, \;\;\;\mathbf{S}\neq0$%
\label{Elko1}
\item[2)] $\sigma\neq0,\;\;\; \omega= 0$\label{dirac1}\qquad5) $\sigma= 0 = \omega, \;\;\;\mathbf{K}=0,\;\;\; \mathbf{S}\neq0$%
\label{tipo4}
\item[3)] $\sigma= 0, \;\;\;\omega\neq0$\label{dirac2} \qquad\!6) $\sigma= 0 = \omega, \;\;\; \mathbf{K}\neq0, \;\;\; \mathbf{S} = 0$%
\end{itemize}
\noindent
The first three classes are composed by Dirac spinor fields and it is implicit that in this case  $\mathbf{J}$\textbf{, }$\mathbf{K}$\textbf{, }$\mathbf{S}$ $\neq0$.
In particular, for a Dirac spinor fields describing an electron, {\bf J} is a future-oriented timelike current vector providing the current of probability; {\bf S} is the distribution of intrinsic angular momentum, and the spacelike vector {\bf K} is associated to the direction of the electron spin.

A Majorana spinor field belongs to the class (5), while Weyl spinor fields are in the class (6). Type-(4) spinor fields are the so-called flag-dipole spinor fields. Furthermore, if $\psi$ is a typical Dirac spinor field  and $\zeta$ is an arbitrary spinor field such that $\zeta\el{\dagger}\gamma_{0}\neq 0$,  $\psi$ is herewith proportional to $Z\zeta$, where $Z$ is given by Eq. (\ref{ZZZ}).

Before delving deeper into the investigation of some interesting outputs in this approach, let us first emphasize that there are no other possible classes for the spinor fields based on different bilinear covariants. In fact, when $\sigma\neq 0$ and/or $\omega\neq 0$, it implies that $\mathbf{S}\neq 0$ and $\mathbf{K}\neq 0$ --- note that $J\el{0}>0$ and hence $\mathbf{J}$ does not equal zero. Besides, the constraint $\omega=0=\sigma$ implies that $Z=\mathbf{J}(1+i(\mathbf{s}+h\gamma_{0123}))$, where $(\mathbf{s}+h\gamma_{0123})\el{2}=-1$, $\mathbf{s}$ is a spacelike vector, and $h$ a real number given by $h=\pm\sqrt{1+\mathbf{s}^2}$. In this vein $\mathbf{J}(\mathbf{s}+h\gamma_{0123})=\mathbf{S}+\mathbf{K}\gamma_{0123}$. 
It is useful to provide further  features of type-(4) spinor fields. For flag-dipole spinor fields, Eq. (\ref{ZZZ}) gives $ Z=\mathbf{J}+i\mathbf{J}s-ih\gamma_{0123}\mathbf{J},$ where $s=\|{\bf s}\|$.  It implies forthwith that $(1+is-ih\gamma_{0123})Z=0$, and taking into account that $\mathbf{J}\el{2}=0$ for type-(4) spinor fields,  $Z$ is shown to be Clifford multivector satisfying  $Z\el{2}=0$. Such spinor fields were widely investigated in \cite{hopf} in a more topological geometric context, as well as some interesting applications.

 The bilinear
covariant $\mathbf{S}$ in (\ref{fierz}) is given by  $\mathbf{S} =  \mathbf{J}\wedge {\bf s}$. For type-(4) spinor fields the real coefficient satisfies $h\neq 0$. Lounesto shows that either ${\bf J}^2=0$ or $(s - ih\gamma_{0123})^2 = -1$. The helicity $h$ relates  \textbf{K}
and \textbf{J} by \textbf{K} = $h$\textbf{J}. The definition of helicity $h$ in terms of bilinear covariants precedes and implies the definition of helicity in quantum mechanics, as well the equivalent relation for anti-particles \cite{boehmergrafeno}. Such approach further prov ides a straightforward form
for the Hamiltonian describing the one-layer superconductor graphene, given by Tr($\gamma^5{\bf K}\gamma^0)$  \cite{boehmergrafeno}.

\section{Peculiar Features}

Roughly speaking, the framework of Lounesto's classification allows a twofold approach: on the one hand it is possible to study and classify new spinor fields recently discovered in the literature. Moreover, their geometric content can be explored and it sheds new light in the investigation on their physical content. We shall deal with this aspect in the following two Subsections. On the another hand, it permits the exploration of genuinely different spinor fields, without any physical counterpart. We delve into this issue in the third Subsection.

\subsection{Elko spinor fields and its properties}

Imagine a mass dimension one spinor field with $1/2$ spin, obeying the Klein-Gordon, but \emph{not the  Dirac} field equations. Endowed with such predicates, it is indeed possible to call that spinor field as {\it strange}. In what follows, however, we shall argue that the strangeness of such spinor, the called Elko spinor, is far from pejorative.

Elko spinor fields are eigenspinors of the charge conjugation operator with eigenvalues $\pm 1$. The plus [minus] sign stands for
{self-conjugate} [{anti self-conjugate}] spinors $\lambda^{S}({\bf p})$ [$\lambda^{A}({\bf
p})]$. 
 Elko spinor fields arise from the equation of helicity
$(\sigma\cdot\widehat{\bf{p}})\phi^{\pm}(\mathbf{0})=\pm
\phi^{\pm}(\mathbf{0})$ \cite{AHL}.   The four
spinor fields are given by
\begin{equation}
\lambda^{S/A}_{\{\mp,\pm \}}({\mathbf
p})=\sqrt{\frac{E+m}{2m}}\Bigg(1\mp \frac{{\bf
p}}{E+m}\Bigg)\lambda^{S/A}_{\{\mp,\pm \}}(\bf{0}),
\label{form}\end{equation}
where $
\lambda^{S/A}_{\{\mp,\pm \}}(\bf{0})=%
\binom{
\pm i \Theta[\phi^{\pm}(\bf{0})]^{*}}{
\phi^{\pm}(\bf{0})}$. The operator $\Theta$ denotes the Wigner's spin-1/2 time reversal operator. 
As 
$\Theta[\phi^{\pm}(\bf{0})]^{*}$ and $\phi^{\pm}(\bf{0})$ present
opposite helicities, Elko cannot be an eigenspinor field of the
helicity operator, and indeed carries both helicities. In order to
guarantee an invariant real norm, as well as positive definite
norm for two Elko spinor fields, and negative definite norm for
the other two, the Elko dual is given by \cite{AHL}
\begin{equation}
\overset{\neg}{\lambda}^{S/A}_{\{\mp,\pm \}}({\bf p})=\pm i \Big[
\lambda^{S/A}_{\{\pm,\mp \}}({\bf p})\Big]^{\dag}\gamma^{0}
\label{dual}.\end{equation}\noindent
 It is useful to
choose $i\Theta=\sigma_{2}$, as in \cite{AHL}, in such a way that
it is possible to express
$
\lambda(\mathbf{p})=\binom{\sigma_{2}\phi_{L}^{\ast}(\mathbf{p})}{\phi_{L}(\mathbf{p})}.
$
The dual is defined in such way that the product $\left({\lambda}^{S/A}_{\{\mp,\pm \}}\right)^\dagger\;\zeta\;{\lambda}^{S/A}_{\{\pm,\mp \}}$ remains invariant under Lorentz transformations. This requirement implies $\zeta=\pm i \gamma\el{0}$ for the Elko case, since it belongs to the $right \oplus left$ representation space \cite{PLBAO}. Endowed with a new dual, Elko respects different orthonormality relations, which engenders non-standard spin sums. Following this reasoning it is possible to envisage the Elko non-locality (see \cite{PLBAO} for the details). Denoting by $\Lambda(\mathbf{x},t)$ the quantum field constructed out of Elko spinor fields as the expansion coefficients and $\Pi(\mathbf{x},t)$ its conjugate momentum, although the following property
\be \{\Lambda(\mathbf{x},t),\Lambda(\mathbf{x'},t)\}=0=\{\Pi(\mathbf{x},t),\Pi(\mathbf{x'},t)\} \ee holds, an unexpected anti-commutation relation is elicited \cite{AHL}:\be \{\Lambda(\mathbf{x},t),\Pi(\mathbf{x'},t)\}=i\int\frac{d\el{3}p}{(2\pi)\el{2}}\frac{1}{2m}e\el{i\mathbf{p}\cdot (\mathbf{x}-\mathbf{x'})}2m[1+G(\mathbf{p})].\label{label} \ee Here $1$ stands for the identity matrix and $G(\mathbf{p})=\gamma^5\gamma_\mu n^\mu$ is a factor arising from the spin sums. The vector $n^\mu = (0,{\bf n})$ defines some preferential direction \cite{AHL}, where ${\bf n} = \frac{1}{\sin\theta}\frac{d\hat{{\bf p}}}{d\phi}$. It was recently demonstrated \cite{WAL}, by explicitly calculation, that the integration over the second term of equation (\ref{label}) equals zero. This is a crucial point, since this term decides the locality structure of the quantum field.

The mass dimension one related to such spinor fields severely suppresses the possible couplings to other fields of the standard model. In fact, if we keep in mind power counting arguments, Elko spinor fields may interact --- in a perturbative renormalizable way --- with itself and with a scalar (Higgs) field. Obviously, the former type of  interaction means an unsuppressed quartic self interaction. At this point it is important to remark that this feature (quartic self interaction) is present in the dark matter characteristics observations \cite{EXP}. Therefore Elko spinor fields seems to perform an adequate fermionic dark matter candidate.

It is worth notice that the appearance of the $G(\bf{p})$ function in the spin sums, however, shall not be underestimated. Its presence turns out to be impossible to conciliate Elko quantum field to the full Lorentz group. Nevertheless, Elko fields are, in fact, a spinor representation under the $SIM(2)$ avatar \cite{JAH} of Very Special Relativity (VSR) \cite{VSR}. The group $SIM(2)$ is the largest possible subgroup of VSR which is necessary to define a quantum theory when parity symmetry is violated. Hence, understanding Elko as a  dark matter prime candidate, it may signalize that in the dark matter sector the Lorentz group may not be the underlying relevant group. Indeed, using the Lounesto framework previously outlined, Elko are classified as type-(5) spinor fields, a generalization of Majorana spinor fields carrying both helicities \cite{RJ}. As mentioned in the Introduction, Lounesto classification goes beyond the standard classification by irreducible representations of the Lorentz group $Spin_{+}(1,3)$. From this perspective, it is quite conceivable that the quantum fields, constructed out from expansion coefficients which do not belong to Lorentz representation, do not respect Lorentz symmetries themselves.

\subsection{The usefulness of topologically exotic terms}

Among an extended inventory of  relevant  new physical possibilities arising from  the use of the non-standard spinor fields, we can branch the role of Elko spinor fields as a detector of exotic spacetime structures \cite{RBJ}.
If the base manifold $M$ upon which the theory is built is simply connected, then the first homotopy group $\pi_{1}(M)$ is well known to be trivial. In this case, supposing that $M$ satisfies the assumptions in the Geroch theorem \cite{GER}, there exists merely one possible spin structure. Consequently, the spin-Dirac operator in the formalism is the standard one. Notwithstanding, when non-trivial topologies on $M$ are regarded, there is a non-trivial line bundle on $M$. The set of line bundles and the set of inequivalent spin structures are labeled by elements of the cohomology group $H\el{1}(M, \mathbb{Z}_{2})$  --- the group of the homomorphisms of $\pi_{1}(M)$ into $\mathbb{Z}_{2}$. In this regard, there are several globally different spin structures arising from the different (and inequivalent) patches of the local coverings. The spin-Dirac operator has in this case an additional term, essentially an 1-form field, that reflects the non-trivial topology. Spinor fields associated to these inequivalent spin structures are called {\it exotic} spinor fields.

Let us make those considerations more precise.  Throughout this Section we denote by $Spin_{1,3}$ and $SO_{1,3}$ the components of such groups connected to the identity, for the sake of conciseness. Given the fundamental map, in fact  a two-fold covering relating the orthonormal
coframe bundle and the spinor bundle\footnote{Let $P_{{SO}_{1,3}}(M)$ denote the orthonormal
coframe bundle, that always exist on spin manifolds.
Sections of $P_{{SO}_{1,3}%
}(M\mathbf{)}$ are orthonormal coframes, and sections of $P_{{Spin}%
_{1,3}}(M\mathbf{)}$ are also orthonormal coframes such that
although two coframes differing by a $2\pi$ rotation are distinct, two
coframes differing by a $4\pi$ rotation are identified.}
$
s: P_{\mathrm{Spin}_{1,3}}(M)\rightarrow P_{\mathrm{SO}_{1,3}%
}(M),$ a spin structure on $M$ is a principal fiber bundle $\pi_{s}: P_{Spin_{1,3}}(M)\rightarrow M$ satisfying: $(i)\;\pi(s(p))=\pi_{s}(p)$ for every point $p$ of $P_{Spin_{1,3}}(M)$, where $\pi$ is the projection of $P_{SO_+({1,3})}(M)$ on $M$, and $(ii)\; s(p\phi)=s(p)Ad_{\phi}$. Here given $\phi\in Spin_{1,3}(M)$, we have $Ad_\phi(\kappa) = \phi\kappa\phi^{-1},$ for all $\kappa\in Cl_{1,3}$.
A spin structure $P:=(P_{Spin_{1,3}}(M),s)$ exists solely when the second Stiefel-Whitney class satisfies specific criteria. To our presentation, however, it is remarkable that if $H\el{1}(M,\mathbb{Z}_{2})$ is not trivial, then the spin structure is not uniquely defined. Two spin structures, say $P$ and $\tilde{P}$, are said to be equivalent if there exists a map $\chi: P\rightarrow \tilde{P}$ compatible with $s$ and $\tilde{s}$; they are said to be inequivalent otherwise. Given an arbitrary spinor field $\psi \in$ sec $P_{Spin_{1,3}}(M)\times \mathbb{C}\el{4}$, where sec means ``section of'', to each element of the non-trivial $H\el{1}(M,\mathbb{Z}_{2})$ one can associate a Dirac operator $\nabla$. This construction provides an one-to-one correspondence between elements of $H\el{1}(M,\mathbb{Z}_{2})$ and inequivalent spin structures (for more details see \cite{thomas, RBJ, GER}).

A crucial difference between the exotic and the standard spinor field is  the action of the Dirac operator on exotic spinor fields.
In a non-trivial topology scenario, the Dirac operator changes by an additional one-form field, which is a manifestation of the non-trivial topology. The exotic structure endows the Dirac operator with an additional term given by $a\el{-1}(x)da(x)$, where $x\in M$ and $d$ denotes the exterior derivative operator. The term $\frac{1}{2i\pi}a\el{-1}(x)da(x)$ is real, closed, and defines an integer C${\check{\rm e}}$ch cohomology class \cite{ISR}. Using the relation between the C${\check{\rm e}}$ch and the de Rham cohomologies, it follows that \be \oint \frac{1}{2i\pi}a\el{-1}(x)da(x) \in \mathbb{Z}.\label{integ} \ee When Dirac spinor fields are regarded, the exotic term can be absorbed into a new shifted potential $A \mapsto A +\frac{1}{2i\pi}a\el{-1}(x)da(x)$: the exotic term may be understood as an external electromagnetic potential that is summed to the physical electromagnetic potential, which
plays the role of a disguise for the exotic term. In this way the
exotic spacetime structures
can not be probed by Dirac spinor fields, which perceive the exotic term as an effective electromagnetic potential.

From the perspective of Elko spinor fields, however, the situation changes drastically. The reason is that the spinor field discussed in the previous Section is an eigenspinor of the charge conjugation operator. Therefore it does not carry local $U(1)$ charge of the standard type. Hence, any type of extra term present in the Dirac operator cannot be absorbed into the electromagnetic potential. As it is extensively discussed in \cite{GER}, the exotic term may be expressed as $\frac{a(x)}{\sqrt{2\pi}}=\exp{(i\theta(x))} \in U(1)$. It yields \ba \frac{1}{2\pi}a\el{-1}(x)da(x)&=&\exp{(-i\theta(x))}(i\gamma\el{\mu}\nabla_{\mu}\theta(x))\exp{(i\theta(x))}\nonumber\\&=&i\gamma\el{\mu}\partial_{\mu}\theta(x).\label{becel}\ea Now, making the conceivable exigency that the exotic Dirac operator must be considered the square root of the Klein-Gordon operator, we have\footnote{Hereon  we are not going to specify the different Elko types, which simplify the content of indexes in Eq. (\ref{udu}). Again, for a complete discussion, see \cite{RBJ}.} \ba [i\gamma^{\mu}(\nabla_{\mu}+\partial_{\mu}\theta)\pm m][i\gamma^{\nu}(\nabla_{\nu}+\partial_{\nu}\theta)\mp m]\lambda\nonumber\\\qquad=(g\el{\mu\nu}\nabla_{\mu}\nabla_{\nu}+m\el{2})\lambda=0.\label{udu} \ea Therefore, the corresponding Klein-Gordon equation for the exotic Elko spinor field reads \be (\Box+m\el{2}+g\el{\mu\nu}\nabla_{\mu}\nabla_{\nu}\theta+\partial\el{\mu}\theta\nabla_{\mu}+\partial\el{\mu}\theta\partial_{\mu}\theta)\lambda=0.\label{udu2} \ee Finally, in order to have the Klein-Gordon propagator for the exotic Elko, as in the standard one, it follows from Eq. (\ref{udu2}) that \be (\Box\theta(x)+\partial\el{\mu}\theta(x)\nabla_{\mu}+\partial\el{\mu}\theta(x)\partial_{\mu}\theta(x))\lambda=0\label{ofim}. \ee
The result encoded in Eq. (\ref{ofim}) makes Elko spinor field a very useful tool to explore unusual topologies in many contexts. Indeed Eq.  (\ref{ofim}) asserts that the Elko spinor structure constrains the exotic term related to the non-trivial spacetime topology. The possibility of  extracting information from the subjacent topology without using any additional (sometimes ill defined) shifted potentials is, in fact, quite attractive. Equation further (\ref{ofim}) encompasses the relationship between gravitational sources induced by exotic topologies. Recently the combined action of a spinor field coupled to the gravitational field was obtained in \cite{mal1}. Furthermore, Eq.(\ref{ofim}) complies with the differential-topological restrictions on the spacetime for the evolution of our Universe. The differential-geometric description of matter by differential structures of spacetime might leads to a unifying model of matter, dark matter and dark energy. Indeed, by taking into account exotic differential structures, it may be the source of the observed anomalies without modifying the Einstein equations or introducing unusual types of matter, as  a vast resource of possible explanations for recently observed surprising astrophysical data at the cosmological scale, merely provided by differential topology \cite{mal1}.

Furthermore, such exoticness induces a dynamical mass which is embedded in the VSR framework \cite{plb1}. It is accomplished  by identifying the VSR preferential direction with a dynamical dependence on the {\em kink} solution of a $\lambda \phi^{4}$ theory, for a scalar field $\phi$. The exotic term $\partial_\mu\theta$ is chosen to be $v_\mu\phi$, where $v_\mu$ provides a preferential direction, an inherent preferred axis --- along which Elko is local.
This is solely one among various possible scenarios, using exotic couplings among dark spinor fields and scalar field topological solutions \cite{plb1}.

 \subsection{The appearance of new spinors}

In the specific context of $f(R)$-cosmology, it was recently reported a solution for the Dirac equation with torsion, considering  Bianchi type-I cosmological models \cite{FABB}. The gravitational dynamics of the theory may be described by the metric and its compatible connection, or alternatively by the tetrad field and the spin-connection as well. The equations of motion are
\ba f'(R)R_{\rho\sigma}-\frac{1}{2}f(R)g_{\rho\sigma}&=&\Sigma_{\rho\sigma}\label{mot1}\nonumber\\ \frac{1}{2}\Bigg(\frac{\partial f'(R)}{\partial x\el{\alpha}}+S_{\alpha\gamma}\el{\;\;\,\gamma}\Bigg)(\delta_{\sigma}\el{\alpha}\delta_{\rho}\el{\beta}-\delta_{\rho}\el{\alpha}\delta_{\sigma}\el{\beta})+S_{\rho\sigma}\el{\;\;\,\beta}&=&f'(R)\,T_{\rho\sigma}\el{\;\;\,\beta}, \nonumber\ea where $R_{\rho\sigma}$ is the Ricci tensor and $T_{\rho\sigma}\el{\;\;\,\beta}$ stands for the torsion tensor. The quantities $\sigma_{\rho\sigma}$ and $S_{\rho\sigma}\el{\;\;\,\alpha}$ are the stress-energy and spin tensors of the matter fields.  The energy-momentum tensor is given by $\Sigma_{\rho\sigma}$.  The idea is to couple $f(R)$-gravity to spinor fields and to a spinless perfect fluid. These spinor fields are shown   \emph{not} to be Dirac spinor fields \cite{TEO}. In addition the second equation of motion assents the existence of torsion even in the absence of spinor fields.
Implementing all the necessary constraints, it is possible to show that the spinor solutions reads
\begin{eqnarray}
\label{restrictedspinor1}
&\psi_{1}=\frac{1}{\sqrt{2\tau}}\left(\begin{tabular}{c}
$\sqrt{A-B}\cos{\zeta_{1}}e^{i\theta_{1}}$\\
$0$\\
$0$\\
$\sqrt{A+B}\sin{\zeta_{2}}e^{i\theta_{2}}$
\end{tabular}\right),\\
&\psi_{2}=\frac{1}{\sqrt{2\tau}}\left(\begin{tabular}{c}
$0$\\
$\sqrt{A+B}\cos{\zeta_{1}}e^{i\vartheta_{1}}$\\
$\sqrt{A-B}\sin{\zeta_{2}}e^{i\vartheta_{2}}$\\
$0$
\end{tabular}\right),
\end{eqnarray}
where $A$ and $B$ are constants, the angular functions have time dependence, and $\tau$ is defined as the product of the scale factors appearing in the Bianchi type-I model (not relevant to our purposes). The point to be stressed is that, after a tedious calculation, the bilinear covariants associated to $\psi_{1}$ and $\psi_{2}$ classify the spinor fields (\ref{restrictedspinor1}) as type-(4):  legitimate flag-dipole spinor fields that are obtained when the Dirac equation with torsion is regarded in the $f(R)$-cosmological scenario \cite{ENO}. It is the first time, up to our knowledge, that a physical solution corresponds to a type-(4) spinor\footnote{This fact is more remarkable than it may sound. Several spinor solutions are of the form presented in (\ref{restrictedspinor1}). Notwithstanding, after all, the class under Lounesto's classification appears to be other than type-(4). For instance, on page $65$ of \cite{LL} it is possible to find similar structured spinor fields. Twenty pages of calculations led the authors to the very exciting conclusion that they belong to the type-(4) set. After some ponderation, however, we were brought back to the Earth: professor Leite Lopes' book was not wrote using the Weyl representation!}.
Eq. (\ref{restrictedspinor1}) evinces a physical manifestation of type-(4), or flag-dipole, spinor fields according to Lounesto's classification.

We finalize this Section by pointing out a provocative interpretation of the type-(4) spinor fields as manifested via Eq. (\ref{restrictedspinor1}). There is no quantum field constructed out yet with type-(4) spinor fields and it is certainly an interesting branch of research. In view of the analysis of Sec. IIIA, such a quantum field shall not respect Lorentz symmetry. From this perspective, it would be the darkest possible candidate to dark matter. Being more conservative, without making any reference to its possible quantum field, type-(4) spinor fields, as it appears, are also quite provocative. Usually, generalizations of General Relativity are studied to give account of cosmological problems, without  appealing to the existence of dark matter, for instance. Nevertheless, as we have mentioned, type-(4) spinor fields appeared only in a, double, generalization of General Relativity. Moreover, the presence of torsion in a $f(R)$ gravity is crucial to the functional form of these spinor fields as explicit in (\ref{restrictedspinor1}). Hence, type-(4) spinor fields, a essentially dark spinor (we restrain to say dark matter), comes up in a far from usual gravitational theory, which is commonly investigated to preclude the necessity of ``dark'' objects.

\section{Final Remarks}

A plethora of open questions still haunts (in particular) theoretical physicists. The non-standard spinor fields --- both under Lounesto as well as Wigner classification ---
are an evidently useful alternative
to pave the road to solve some questions, mainly in field theory and cosmology/gravitation.
It brings some nice and unexpected properties, like the existence of fermions with mass dimension one and a subtle Lorentz symmetry breaking, for instance.
Facing such paradigm shift seems to
upheaval what we know already about
field theory and the elementary particles description, which were restricted to Dirac, Majorana and Weyl spinor fields heretofore, in Minkowski spacetime.
As we have shown, flag-dipole type-(4) spinor fields are physical solutions of the Dirac equation with torsion in the context of $f(R)$-cosmology. Furthermore, Elko spinor fields representing type-(5), abreast of  Majorana spinor fields,  are evinced to be prime candidates to describe dark matter. We moreover have introduced the exotic dark spinor fields, which dynamics constraints  both the spacetime metric structure and the non-trivial topology of the universe. In particular, it brings exotic couplings among dark spinor fields and scalar field topological solutions. The topics here introduced are merely the tip of the iceberg, and
there are more useful properties on spinor fields (and their application in physics) still to be explored. \section*{Acknowledgments}

The authors would like to thanks Prof. Jos\'e  Abdalla Helay\"el-Neto for the continuous motivation.
R. da Rocha is grateful to Conselho Nacional de Desenvolvimento Cient\'{\i}fico e Tecnol\'ogico (CNPq)  grants 476580/2010-2 and 304862/2009-6 for financial support. J. M. Hoff da Silva thanks to CNPq (482043/2011-3) for partial support.

\end{document}